\documentclass[letterpaper, 10 pt, conference]{ieeeconf}

\IEEEoverridecommandlockouts                              

\usepackage{graphicx}      
\usepackage{amssymb,amsmath}
\usepackage{color,soul}
\usepackage{subfigure}
\usepackage{epstopdf}   

\newtheorem{lem}{Lemma}

\newtheorem{pro}{Program}

\usepackage{url}

\usepackage{boldline} 

\DeclareMathOperator*{\argminA}{arg\,min}
\newcommand{\norm}[1]{\left\lVert#1\right\rVert}

\title{\LARGE \bf
Structured exploration in the finite horizon \\ linear quadratic dual control problem}
\author{Andrea Iannelli$^{1}$, Mohammad Khosravi$^{1}$  and Roy S. Smith$^{1}$
\thanks{$^{1}$ The authors are with the Department of Information Technology and Electrical Engineering, Automatic Control Lab, ETH, Z\"{u}rich 8092, Switzerland
{\tt\small iannelli/khosravm/rsmith@control.ee.ethz.ch}}
}

\begin{document}

\maketitle
\thispagestyle{empty}
\pagestyle{empty}

\begin{abstract}                
This paper presents a novel approach to synthesize dual controllers for unknown linear time-invariant systems with the tasks of optimizing a quadratic cost while reducing the uncertainty.
To this end,
a synthesis problem is defined where the feedback law has to simultaneously gain knowledge of the system and robustly optimize the cost.
By framing the problem in a finite horizon setting, the trade-offs arising when the tasks include both identification and control are formally captured in the optimization problem. Results show
that efficient exploration strategies are achieved when the structure of the problem is exploited.
\end{abstract}

\section{Introduction}
One of the most widespread approaches in control is the Linear Quadratic (LQ) regulator, whereby the goal is to design a feedback law which minimizes deviations of the states from a desired reference trajectory (e.g. the origin) while keeping as small as possible the necessary action. In the full state information case (standard LQR), when the dynamics is \emph{exactly} known the problem has a well known optimal solution \cite{Berts_DP_OC}. In the infinite horizon case, that is when transient features are negligible and the problem is approximated in an infinitely long time window, this consists of a static gain matrix associated with the solution of an Algebraic Riccati Equation (ARE). When the problem is studied on a finite horizon, the optimal feedback law is time-varying and is associated with a Riccati difference (or differential) equation (DRDE).

Despite important control theoretic works on robust $\mathcal{H}_2$ analysis and filtering problems \cite{Sznaier_robH2,Sun_Packard_H2LFT}, the solution of the LQ control problem when the dynamics is unknown is far less understood. 
Notably, this has been used in the last few years as case study to show possible complementarities of Reinforcement Learning (RL) and control theory-based approaches for the fundamental problem of optimally manipulating an unknown system by using the information carried in the collected data \cite{Recht_tour}.
Bridging these two communities has been the effort of many recent works, see e.g. \cite{Cohen_onlineLQR,Dean2019,Matni2019}, but despite the variety of techniques considered, no strategies which allow for an easy implementation on one hand, and provide optimal cost guarantees on the other, have been found \cite{Recht_tour}.
Moreover, a fundamental unsolved problem is what is the best strategy to extract information about the system such that the performance can be improved while preserving at the same time safety. In other words, borrowing terminology from the reinforcement learning community (e.g. multi-armed bandits, \cite{Sutton1998}), specifying an optimal \emph{policy} (control law) that robustly balances \emph{exploration} (acquiring knowledge of the system by testing and identification) and \emph{exploitation} (operating the system to maximize the \emph{reward}, or performance).

The approach considered here owes to the long and rich tradition of dual control \cite{Astrom_dual}, where the problem of simultaneously identifying and controlling a system was first formalized, and experiment design, whereby one attempts to determine the most suitable inputs in order to extract information from the unknown plant \cite{ExpDes_CSM17}.
The material presented here is also inspired by a recent publication (\cite{Ferizbegovic_2020}) where the unknown LQ problem is framed in a dual control setting. Specifically, given an initial estimate of the dynamics in the form of nominal state matrices and an ellipsoidal uncertainty set, the joint optimization of two robust feedback laws $G_{\text{K1}}$ and $G_{\text{K2}}$ is proposed therein considering two distinct infinite horizon problems.

While retaining the same application-oriented philosophy, i.e. promoting a reduction of the uncertainty which is beneficial for the purpose of minimizing a given cost,
the work here substantially changes the synthesis approach by framing the problem in a finite horizon setting. This is motivated by the fact that the dual control problem is more realistically described in a finite time window, due to the importance of transient features.
The design of two robust static feedback laws tasked with different goals is thus shifted to that of a single time-varying law $G_{\text{K}}$, responsible for dealing simultaneously with
the two tasks. From an optimal control perspective, but in a dual control setting now, the problem is formulated as the solution of an DRDE rather than of an ARE. The main advantage is that by framing the problem in a finite horizon setting the different trade-offs between exploration and exploitation are captured and can be optimized over. New insights into these trade-offs are in turn believed to
help gaining a deeper understanding of the unknown LQ problem.

The main technical contribution of the paper is the formulation of a Semidefinite program (SDP) to solve the robust dual LQR control problem in the finite horizon setting optimally balancing exploration and exploitation. This is presented in Section \ref{SDP_formulation}, where also the corresponding programs for the nominal and robust (but with fixed uncertainty, i.e. without exploration) problems are derived. The other important contribution is gathered in Section \ref{Results}, where features of the synthesised policies are shown through numerical examples. The application-oriented nature of the exploration strategy proposed here is termed \emph{structured}, in order to emphasize the ability to exploit specific properties of the analysed system. 

\section{Problem description}\label{Background}

\subsection{Background}
Consider the discrete linear time-invariant system:
\begin{equation}\label{system}
x_{t+1}=A x_{t}+B u_{t}+w_{t},\quad w_{t}\sim \mathcal{N}(0, \sigma_w^2 I_{n_x}),\quad x_0=0,  
\end{equation}
where $x_{t} \in \mathbb{R}^{n_x}$ is the (measured) state, $u_{t} \in \mathbb{R}^{n_u}$ is the control input, and  $w_{t} \in \mathbb{R}^{n_x}$ is the normally distributed process noise with zero mean and covariance $\sigma_w^2 I_{n_x}$. Given cost matrices $Q \succeq 0$ and $R \succeq 0$, the objective is to design a feedback law minimizing the expected finite horizon quadratic cost $J$ in $[1,T]$ (with $1 < T < \infty$):
\begin{equation}\label{fh_cost}
J= \mathbb{E} \left[ \sum_{t=1}^{T-1}\left( x_{t}^{\top} Q x_{t}+u_{t}^{\top} R u_{t}\right)+ x_{T}^{\top} Q x_{T} \right],
\end{equation}
where the expectation is with respect to $w$. When $A$ and $B$ are known, the optimal input is given by
the time-varying state-feedback law $u_t=K_t^{\text{DRDE}} x_t$, where $K_t^{\text{DRDE}}$ is associated with the stabilizing solution of the discrete time Riccati difference equation (DRDE) for (\ref{system}).

The case of unknown $A$ and $B$, where the only access to information on (\ref{system}) is through measurements of $x$ and $u$, is considered here. 
An estimation of the unknown dynamics is obtained through the so-called Coarse-ID approach (\cite{Dean2019}).
Given a dataset made of $N$  samples $\mathcal{S}=\{(x_t, u_t): 1\leq t \leq N \}$, the \emph{nominal dynamics} is estimated through the least squares problem:
\begin{equation}\label{unc_model_Nom}
(\hat{A},\hat{B})= \argminA_{A,B} \sum_{t=1}^{N-1} \norm{-x_{t+1}+ A x_{t}+B u_{t} }^2_2, 
\end{equation}
and the \emph{true dynamics} belongs to the ellipsoidal set $\Omega$: 
%
\begin{subequations}\label{unc_model}
\begin{align}
 \Omega(X,D)&=\{X: X^\top D X \preceq I  \}, \quad D \in \mathbb{S}^{n_x+n_u}, \label{unc_model_1}\\
X&= \begin{bmatrix}
( \hat{A}-A )^\top   \\
 ( \hat{B}-B )^\top   \\
\end{bmatrix}, \quad X \in \mathbb{R}^{(n_x+n_u)\times n_x}, \label{unc_model_2}
\end{align}
\end{subequations}
where $D$ defines the uncertainty. A possible estimate for $D$ can be obtained by making use of an empirical Bayes argument and taking it as the variance of the posterior distribution of $(A,B)$ given $\mathcal{S}$ (\cite{Umenberger_RRL19}). Precisely, (\ref{unc_model_1}) holds with probability 1-$\delta$ for:
\begin{equation}\label{unc_model_D}
 D= \frac{1}{c_{\chi} \sigma_w^2}  \sum_{t=1}^{N-1} \begin{bmatrix}
x_{t}   \\
 u_{t}   \\
\end{bmatrix} \begin{bmatrix}
x_{t}   \\
 u_{t}   \\
\end{bmatrix}^{\top}, \quad c_{\chi} = \chi^2_{(n_x^2+n_x n_u),\delta},
\end{equation}
where $\chi^2_{n,\delta}$ 
is the critical value for a Chi-squared distribution with $n$ degrees of freedom and probability level $\delta$.

Given this uncertainty description, the objective is to synthesize a policy $G_{\text{K}}$ that minimizes the worst-case $J$:
\begin{equation}\label{fh_cost_WC}
J_{\text{WC}} = \min_{G_{\text{K}}} \max_{(A,B) \in \Omega(X,D)} J.
\end{equation}
To this end, $G_{\text{K}}(K_t,S_{t})$ is parametrized as:
\begin{equation}\label{feedback_law}
u_{t}=K_t x_{t}+e_{t},\quad \quad e_{t}\sim \mathcal{N}(0, S_{t}),\quad S_{t} \in \mathbb{S}^{n_u}.
\end{equation}
The law consists of a time-varying state-feedback part, and a random excitation input $e_t$ with time-varying covariance for the purpose of exploration.
The time-varying formulation of $S_{t}$
captures the fact that, as knowledge of the system is acquired, the random part of the excitation should decrease.
This aspect is also found in several methods proposed in the RL community, e.g. the concept of exploration rate in $\epsilon$-greedy algorithms \cite{Sutton1998}.
Formal ways for adapting rates to the learning progress have been proposed in the learning literature \cite{Auer2002,Boltzmann_Exploration}, e.g. leveraging the concept of regret bounds \cite{Matni2019}. In this formulation, $S_{t}$ will be an optimization variable and thus its value will depend,
among other things, on the properties of the system to be identified.


\subsection{Motivating example}
The importance of formulating the dual control problem in a finite horizon, rather than in an infinite one as proposed in \cite{Ferizbegovic_2020} and in general in the recent learning-LQR literature \cite{Cohen_onlineLQR,Dean2019}, is discussed here.

Consider the scenario where the unknown system (\ref{system}) has to be operated over a certain horizon $[1,T]$ while minimizing (\ref{fh_cost}). Since the dynamics are not known, a simple strategy consists of choosing an intermediate time $T_{sw}<T$, and dividing the horizon into two phases. In the first (exploration, or \emph{ID-phase}), the system is excited with random input $u_t\sim \mathcal{N}(0, \sigma_u^2 I)$ and the measured response (e.g. in the form of $\mathcal{S}$) is used to identify the nominal matrices through (\ref{unc_model_Nom}). In the second (exploitation, or \emph{K-phase}), a controller which optimizes (\ref{fh_cost}) for the identified nominal matrices is synthesised. One possible option is the use of time-varying feedback $K_t^{\text{DRDE}}$ on the remaining horizon $[T_{sw},T]$.
That is, a pure exploration phase is followed by a pure exploitation phase (often termed \emph{explore-then-commit} in the RL literature \cite{Garivier_ExpCommit}).
This clearly leads to a trade-off between the \emph{duration} of these two phases, where for high $T_{sw}$ the benefit of estimating the model more accurately contrasts with the disadvantage of optimally controlling the plant for a shorter time, and viceversa. 

\begin{figure}[h!]%
\begin{center}
\includegraphics[width=0.8\columnwidth]{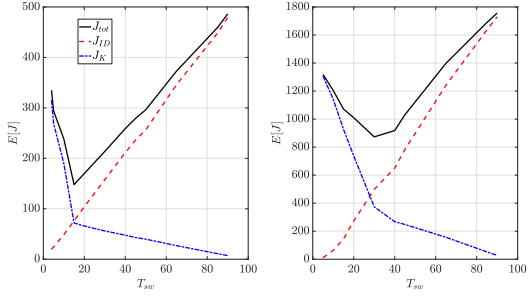}
\caption{Expected costs for the two unknown plants as a function of the switching time $T_{sw}$.}
\label{FHqualitative}
\end{center}
\end{figure}

In order to exemplify this aspect, an experiment is performed on a horizon of length $T$=100 with two randomly generated stable plants having $n_x$=3, $n_u$=2, and using $\sigma_u$=1, $\sigma_w$=0.5. Figure \ref{FHqualitative} shows the total expected cost $J_{\text{tot}}$ (obtained by averaging over 100 realizations of noise and random excitation) as a function of $T_{sw}$.
The total cost can be broken down into $J_{\text{ID}}$ and $J_{\text{K}}$, associated respectively with the identification part in the horizon $[1, T_{sw}]$ and with the deployment of the controller in $[T_{sw},T]$.

It can be observed that there exists an \emph{optimal} switching time where the benefit of further exploring the unknown dynamics is overcome by the cost of exploration. 
This trade-off depends on the unknown true system, and
it can only be captured in a finite horizon setting, where distinctive transient features of the \emph{ID-phase} and \emph{K-phase} are retained.
While the explore-then-commit strategy tested here captures fundamental conflicting aspects arising in the dual control problem, it has important limitations, among which are robustness ($K_t^{\text{DRDE}}$ is not robust to estimation errors) and optimality (the interplay between identification and control is not exploited since exploration and exploitation are sequentially applied). 
The next section addresses these fundamental aspects of the dual control problem by proposing a novel synthesis strategy.

\section{Semidefinite programs formulations}\label{SDP_formulation}

The goal is to derive a convex formulation for synthesising a feedback law $G_{\text{K}}$ (\ref{feedback_law}) that optimizes $J_{\text{WC}}$ (\ref{fh_cost_WC}). In order to clearly present the steps involved and highlight their meaning, the presentation is broken down into 3 parts. Section \ref{SDP_NomDes} deals with the nominal case (where the estimated system coincides with the true one, i.e. $A=\hat{A}$ and $B=\hat{B}$), conceptually equivalent to solving the associated DRDE. Section \ref{SDP_Rob} considers a worst-case design where the set of uncertainty is fixed throughout the horizon, which thus is a robust version of the DRDE. Finally, Section \ref{SDP_Dual} establishes the dual control formulation, where exploration is promoted and thus the uncertainty of the system can be reduced while robustly controlling the plant.

The key idea is to use the application-oriented approach first introduced in \cite{Ferizbegovic_2020},
with the important differences that the problem is formulated in the finite horizon and a single (time-varying) policy responsible for simultaneously exploring and controlling is synthesized.
Exploration is used only to update the uncertainty matrix $D$ (\ref{unc_model_D}), while the nominal matrices $\hat{A}$ and $\hat{B}$ are kept fixed.
This is in line with related works \cite{Ferizbegovic_2020,Umenberger_RRL19} that 
make use of the certainty equivalence assumption.

The first step consists of deriving an expression for the cost $J$ which can be used in the robust optimization problem (\ref{fh_cost_WC}). Let us begin by denoting by $P_t$ the
covariance matrix of the state
at timestep $t$:
\begin{equation}\label{cov_def}
P_t=\mathbb{E} \left [
x_{t}
x_{t}^{\top} \right]  \in \mathbb{S}^{n_x}.
\end{equation}
Define also $\bar{Q}_t:=\left[ \begin{smallmatrix}Q^{\frac{1}{2}} \\ R^{\frac{1}{2}}K_t \end{smallmatrix}\right]\in \mathbb{R}^{(n_x+n_u) \times n_x}$, $\bar{R}:=\left[ \begin{smallmatrix}0 \\ R^{\frac{1}{2}} \end{smallmatrix}\right]\in \mathbb{R}^{(n_x+n_u) \times n_x}$. Then the following result, proved in the Appendix, holds.
\begin{lem}\label{LemmaCost}
The cost $J$ in (\ref{fh_cost}) is equivalent to:
\begin{equation}\label{fh_cost_Pi}
J= \textnormal{Tr} \left(\sum_{t=1}^{T-1} \left( \bar{Q}_t P_t \bar{Q}_t^{\top}+ \bar{R} S_t \bar{R}^{\top} \right)+Q P_T Q^{\top}\right).
\end{equation}
\end{lem}
The benefit of Lemma \ref{LemmaCost} is that it allows the finite horizon cost to be rewritten as a function of the covariances $P_t$, as it is the case for the infinite horizon cost, whose minimization in turn is equivalent to the computation of the $\mathcal{H}_2$ norm of (\ref{system}).

\subsection{Nominal design}\label{SDP_NomDes}
The solution to the nominal problem, as in the SDP-based computation of the $\mathcal{H}_2$ norm, can be obtained by
minimizing (\ref{fh_cost_Pi}) while constraining the covariance $P_t$ to satisfy $\forall t \in [1,T-1]$ the discrete time Lyapunov inequalities associated with the closed loop (\ref{system})-(\ref{feedback_law}):
\begin{subequations}\label{SDP_nom_base}
\begin{align}
&\min_{G_{\text{K}}} \hspace{.1in} \textnormal{Tr} \left(\sum_{t=1}^{T-1} \left( \bar{Q}_t P_t \bar{Q}_t^{\top}+ \bar{R} S_t \bar{R}^{\top} \right)+Q P_T Q^{\top}\right), \label{SDP_nom_base_1}\\
&P_{t+1}\succeq (A+B K_t)P_t(A+B K_t)^{\top}+\sigma^2_{w}I+B S_t B^{\top}, \label{SDP_nom_base_2}\\
&P_1\succeq \sigma^2_{w}I,\hspace{.15in} \label{SDP_nom_base_3}
\end{align}
\end{subequations}
where (\ref{SDP_nom_base_3}) comes from the assumed zero initial condition on the state (that is, $P_0\equiv0$).
For generality, and to make more clear the differences among the 3 synthesis approaches, the generic policy $G_{\text{K}}(K_t,S_{t})$ is considered. However, as intuitive and confirmed later by the results, the random excitation part $S_t$ will be zero in this case.
\newline
The program in (\ref{SDP_nom_base}) is convex and can be recast as an SDP with well known Linear Matrix Inequalities (LMI) manipulations \cite{Scherer_notes}. First, (\ref{SDP_nom_base_1}) is upper bounded by replacing the argument of the summation in (\ref{SDP_nom_base_1}) with $Y_t \in \mathbb{S}^{n_x+n_u} \succeq 0$, and the new objective function (\ref{SDP_nom_base_1}) is written by using Schur complement as:
\begin{equation}\label{SDP_nom_obj}
\begin{aligned}
&\min_{G_{\text{K}}} \textnormal{Tr} \left(\sum_{t=1}^{T-1} Y_t+Q P_T Q^{\top}\right),\\
& \begin{bmatrix}
Y_t - \bar{R} S_t \bar{R}^{\top} & \bar{Q}_t P_t  \\
  P_t \bar{Q}_t^{\top} & P_t \\
\end{bmatrix}\succeq 0, \quad \forall t \in [1,T-1]. \\ %
\end{aligned}
\end{equation}
Note that $\bar{Q}_t$ and $P_t$ give rise to bilinear terms, thus the auxiliary variable $Z_t=P_t K_t^{\top}$ is defined.
As for the inequalities (\ref{SDP_nom_base_2}), they can be recast as coupled LMIs by application of Schur complement. This leads to:
\begin{pro}\label{SDP_Nom_pro}Nominal design \ \\
\begin{subequations}\label{SDP_nom_fin}
\begin{align}
\min_{G_{\text{K}}} J=&\min_{Y_t,P_t,Z_t,S_t} \textnormal{Tr} \left(\sum_{t=1}^{T-1} Y_t+Q P_T Q^{\top}\right),\\
& \begin{bmatrix}
Y_t - \bar{R} S_t \bar{R}^{\top} & \left[ \begin{smallmatrix}Q^{\frac{1}{2}}P_t \\ R^{\frac{1}{2}}Z_t^{\top} \end{smallmatrix} \right]   \\
  * & P_t \\
\end{bmatrix}\succeq 0,  \label{SDP_nom_fin_1}\\
& \begin{bmatrix}
P_t & F_t  \\
  * & P_{t+1}-\sigma^2_{w}I - B S_t B^{\top}\\
\end{bmatrix}  \succeq 0,  \label{SDP_nom_fin_2}\\
& Y_t \succeq 0, S_t \succeq 0, P_{t+1}\succeq 0,\hspace{.15in} \forall t \in [1,T-1],  \nonumber \\
& P_1\succeq \sigma^2_{w}I, \nonumber
\end{align}
\end{subequations}
\end{pro}
where $F_t := P_t A^{\top} + Z_t B^{\top}$. Solving Program \ref{SDP_Nom_pro} is conceptually equivalent to solving the DRDE and leads to identical results (see Fig. \ref{Cmprs_Nom}). The advantage of this formulation is that it allows robustness constraints to be enforced and the effect of exploration to be included. 

\subsection{Robust control design}\label{SDP_Rob}

In the unknown dynamics case, the only knowledge is that $(A,B) \in \Omega(X,D_0)$, where $\hat{A}$, $\hat{B}$, and $D_0$ are assumed to be available from prior experiments or approximate knowledge of the system.
Therefore, the LMIs (\ref{SDP_nom_fin_2}) have to be hold for all possible $(A,B)$ inside $\Omega$.
To solve this robust optimization problem, $A$ and $B$ are written as a function of $X$, $\hat{A}$, and $\hat{B}$ using definition (\ref{unc_model_2}) and a Schur complement is applied to overcome the nonlinearity arising from $B S_t B^{\top}$. Then, since $X$ has to lie inside an ellipsoidal set (\ref{unc_model_1}), an extension of the S-lemma to the matrix case, proposed in \cite{Luo_SIAM_opt}, is employed and the following program is proposed.
\begin{pro}\label{SDP_Rob_pro} Robust control design \ \\
\begin{subequations}\label{SDP_Rob_fin}
\begin{align}
J_{\text{WC}}=&\min_{Y_t,P_t,Z_t,S_t,p_t} \textnormal{Tr} \left(\sum_{t=1}^{T-1} Y_t+Q P_T Q^{\top}\right),\\
& \begin{bmatrix}
Y_t - \bar{R} S_t \bar{R}^{\top} &  \left[\begin{smallmatrix}Q^{\frac{1}{2}}P_t \\ R^{\frac{1}{2}}Z_t^{\top} \end{smallmatrix} \right]  \\
  * & P_t \\
\end{bmatrix}\succeq 0,  \label{SDP_Rob_fin_1}\\
& \begin{bmatrix}
\left[ \begin{smallmatrix}P_t & 0\\ * & S_t \end{smallmatrix} \right] & H_t & G_t  \\
 *  & P_{t+1}-\sigma^2_{w}I-p_t I & 0 \\
  * & * & p_t D_0 \\
\end{bmatrix}  \succeq 0, \label{SDP_Rob_fin_2}\\
& Y_t \succeq 0, S_t \succeq 0, P_{t+1}\succeq 0, p_t\geq0, \forall t \in [1,T-1],  \nonumber \\
& P_1\succeq \sigma^2_{w}I, \nonumber
\end{align}
\end{subequations}
\end{pro}
%
%
where $G_t$ := -$\left[ \begin{smallmatrix}P_t & Z_t\\ 0 & S_t \end{smallmatrix} \right]$, $H_t$ := -$G_t \left[ \begin{smallmatrix} \hat{A}^{\top}\\ \hat{B}^{\top} \end{smallmatrix} \right]$,  and $p_t$ is a multiplier from the S-lemma.
The crucial feature of Program \ref{SDP_Rob_pro} is that the ellipsoid $\Omega(X,D_0)$ defining the uncertainty is fixed throughout the horizon.
The consequence of this is that exploration is not encouraged, since it has an associated cost for which is not rewarded. In other words, the generation of control inputs with a different goal than just minimizing the performance objective will inevitably incur in a higher cost (or \emph{regret}).
This argument has a clear interpretation for $S_t$, where the LMIs (\ref{SDP_nom_fin_1})-(\ref{SDP_Rob_fin_1}) show that non-zero $S_t$ always determine an additional contribution to the cost via $Y_t$.
In fact,
the random excitation part $S_t$ will be zero here (as it was commented on earlier for the nominal case).
As for $K_t$, this will
correspond to the stabilizing solution of the DRDE formed by taking at each time-step the worst-case matrices (which are in principle time-varying), i.e. it is the robust optimal policy.

\subsection{Robust dual control design}\label{SDP_Dual}
In order to promote exploration, it is necessary to describe how the feedback law contributes to obtain knowledge of the system. More formally, a mapping between $G_{\text{K}}$ and the uncertainty $D_t$ at a given time $t$ has to be formulated. From its definition in (\ref{unc_model}), the following is proposed:
\begin{equation}\label{unc_model_D_LMI}
D_t = \frac{1}{c_{\chi} \sigma_w^2} \sum_{l=1}^{t}  \begin{bmatrix}
 P_{l} &   Z_{l}  \\
 * &  \left( Z_{l}^{\top} P_{l}^{-1} Z_{l}\right)+ S_l  \\
\end{bmatrix}.
\end{equation}
Note the explicit influence of $S_t$ and $K_t$ (via $Z_t=P_t K_t^{\top}$) on $D_t$, with the policy also having an indirect effect on $P_t$.

Due to the nonlinearity involving $Z_{l}$ and $P_{l}$ in the lower diagonal block, (\ref{unc_model_D_LMI}) cannot be readily used and thus a convex relaxation is sought. To this end, the matrix inequality in (\cite{Ferizbegovic_2020}, Lemma 1) is employed here to formulate, for a given matrix $\bar{K}\in \mathbb{R}^{n_u\times n_x}$, the following lower bound on $D_t$:
\begin{equation}\label{unc_model_D_LMI_cvx}
D_t \succeq \hat{D}_t = \frac{1}{c_{\chi} \sigma_w^2}   \sum_{l=1}^{t}\begin{bmatrix}
 P_{l} &  Z_{l}  \\
 * & Z_{l}^{\top}\bar{K}^{\top}+\bar{K} Z_{l}-\bar{K} P_{l} \bar{K}^{\top}+S_l  \\
\end{bmatrix}.
\end{equation}
The bound is tight when $\bar{K}_t=K_t$. In this work $\bar{K}_{l}$ is chosen as the nominal controller from Program \ref{SDP_Nom_pro}.
The following dual control design problem is then proposed. 

\begin{pro}\label{SDP_Dual_pro} Robust dual control design \ \\
\begin{subequations}\label{SDP_Dual_fin}
\begin{align}
J_{\text{WC}}=&\min_{Y_t,P_t,Z_t,S_t,p_t} \textnormal{Tr} \left(\sum_{t=1}^{T-1} Y_t+Q P_T Q^{\top}\right),\\
& \begin{bmatrix}
Y_t - \bar{R} S_t \bar{R}^{\top} & \left[ \begin{smallmatrix}Q^{\frac{1}{2}}P_t \\ R^{\frac{1}{2}}Z_t^{\top} \end{smallmatrix}  \right] \\
  * & P_t \\
\end{bmatrix}\succeq 0,  \label{SDP_Dual_fin_1}\\
& \begin{bmatrix}
\left[\begin{smallmatrix}P_t & 0\\ * & S_t \end{smallmatrix}\right] & H_t & G_t  \\
 *  & P_{t+1}-\sigma^2_{w}I-p_tI & 0 \\
  * & * & p_t (D_0+\hat{D}_t) \\
\end{bmatrix}  \succeq 0, \label{SDP_Dual_fin_2}\\
& Y_t \succeq 0, S_t \succeq 0,P_{t+1}\succeq 0, p_t \geq0, \forall t \in [1,T], \nonumber \\
& P_1\succeq \sigma^2_{w}I. \nonumber
\end{align}
\end{subequations}
\end{pro}
The bilinearity between $p_t$ and $\hat{D}_t$ is overcome by using a line search on $p_t$ (assumed constant for simplicity, but allowed to be time-varying).

Program \ref{SDP_Dual_pro} clearly shows that the policy $G_{\text{K}}$ can now perform application-oriented exploration. The key enabler is the
policy-dependent and time-varying upper bound on the true uncertainty $\hat{D}_t$
in (\ref{SDP_Dual_fin_2}). The feedback law is indeed optimized so that the system's response will allow the worst-case matrices $A$ and $B$ to be eliminated from the uncertainty set, to the benefit of the feasibility of the LMIs (\ref{SDP_Dual_fin_2}) and in turn of the achievable cost.
Exploration itself, however, has a cost and thus trade-offs will arise. The cost associated with $S_t$ is seen directly in (\ref{SDP_Dual_fin_1}), while that related to $K_t$ can be interpreted as due to the deviation of $K_t$ from the robust optimal policy. The first trade-off is on which part of the policy $G_{\text{K}}(K_t,S_{t})$ should be used for exploration, whether the state-feedback, the random excitation or both. Another trade-off is on which portion of the horizon exploration should be pursued (reminiscent of the scenario in Figure \ref{FHqualitative}). It is important in this regard to note that a conceptually similar (convex) formulation for the mapping (\ref{unc_model_D_LMI_cvx}) between the policy and the uncertainty $D_t$ was proposed in \cite{Ferizbegovic_2020}. However, while here the cost to pay to keep adding contributions to $\hat{D}_t$ is well captured in Program \ref{SDP_Dual_pro}, it is not clear how this can be accounted for in an infinite horizon setting, where $J$ is effectively an averaged cost and thus does not depend on how many terms (i.e. samples) are featured in the summation leading to $\hat{D}_t$.

\section{Numerical examples}\label{Results}

\begin{figure}[h!]%
\begin{center}
\includegraphics[width=1\columnwidth]{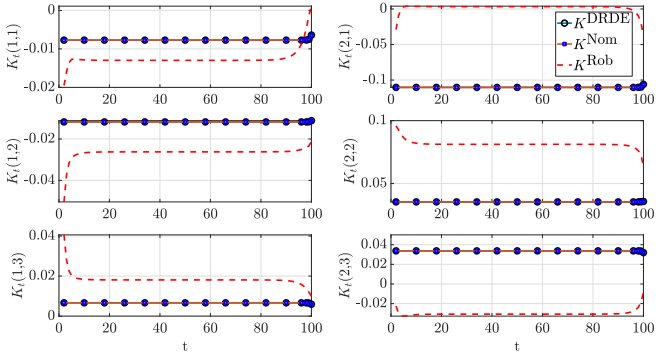}
\caption{Optimal controllers for the nominal and robust (fixed uncertainty) problem.}
\label{Cmprs_Nom}
\end{center}
\vspace{-0.1in}
\end{figure}

Consider the following system:
\begin{equation}\label{Nom1}
\begin{array}{l}
A=\left[
\begin{array}{ccc}
  0.18  &  0.1   &      0\\
         0  &  0.18   & 0.04\\
         0 &  -0.04 &   0.16\\
\end{array}
\right]
\end{array},\quad B=\left[
\begin{array}{cc}
  0  &  1   \\
         0.6  &  0  \\
         0 &  0.6 \\
\end{array}
\right].
\end{equation}
with cost matrices $Q=I_{n_x}$ and $R$=blkdiag(10,1), $\sigma_w^2=0.5$, and $\delta=0.05$. First, a Coarse-ID estimate of the system
%
%
is obtained through 100 simulated roll outs (each of length $T_r=5$) with $u_{t}\sim \mathcal{N}(0, \sigma_u^2I_{n_x})$, $\sigma_u^2=1$. Figure \ref{Cmprs_Nom} shows the time-varying gains of the feedback matrix $K_t$ optimized on the horizon  $[1,100]$ using different design schemes: $K^{\text{DRDE}}$ by solving the DRDE associated with $(\hat{A},\hat{B})$ (\ref{unc_model_Nom}); $K^{\text{Nom}}$ by solving Program \ref{SDP_Nom_pro} for $(\hat{A},\hat{B})$; $K^{\text{Rob}}$ by solving Program \ref{SDP_Rob_pro} for $(\hat{A},\hat{B},D_0)$. The SDP programs are solved using YALMIP \cite{Yalmip} in conjunction with the solver SDPT3 \cite{sdpt3}.

The first observation is that, as expected, $K^{\text{DRDE}}$ coincides with $K^{\text{Nom}}$. Moreover $J_{\text{DRDE}}\cong J_{\text{Nom}}\cong 80$, where the former was obtained from the known closed form solution $J_{\text{DRDE}}=\sum_{t=1}^{T-1}\mathbb{E}\left( w_t^{\top} X_{t+1} w_t \right)$ (where $X_{t}$ is the stabilizing solution of the DRDE, \cite{Berts_DP_OC}), while $J_{\text{Nom}}$ was directly provided by Program \ref{SDP_Nom_pro}. As for the robust design, which achieved $J_{\text{Rob}}\cong 240$, note that $K^{\text{Rob}}$ is generally far from the optimal controller for the nominal plant because of the requirement to guarantee robustness (at the expense of nominal performance).

\begin{figure}[h!]%
\begin{center}
\includegraphics[width=1\columnwidth]{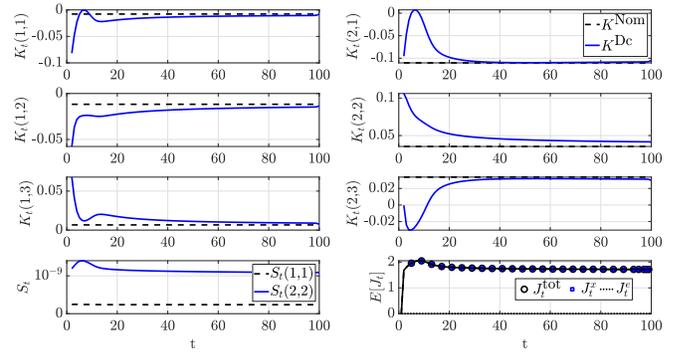}
\caption{Dual controller and cost for system (\ref{Nom1}).}
\label{Rob_Expl_1}
\end{center}
\end{figure}

The dual control policy, designed using Program \ref{SDP_Dual_pro}, is illustrated in Figure \ref{Rob_Expl_1} by comparing the state-feedback dual controller $K^{\text{Dc}}$ with the nominal $K^{\text{Nom}}$, and also by reporting the covariance $S_t$ of the excitation input. The \emph{timestep} cost $J_t^{\text{tot}}$, together with its two contributions $J_t^{x}=\mathbb{E} \left[x_{t}^{\top} \left(Q+K_t^{\top} R K_t \right)x_{t}\right]$ and $J_t^{e}=\mathbb{E}\left[e_{t}^{\top}R e_{t}\right]$, is finally shown in the bottom right plot.

There is a clear exploration action taking place in the first part of the finite horizon, performed only by the state-feedback $K_t$, while the covariance $S_{t}$ is practically zero. This can also be appreciated from the plot with the costs where $J_t^{e} \cong 0$, $J_t^{\text{tot}}$ $=$ $J_t^{x}$, and the latter has an initially increasing and later decreasing trend, before achieving a constant value. Indeed, since the cost would increase linearly in the optimal finite horizon problem, this can be read as a qualitative indication that, after approximately 20 timesteps, $K_t$ has stopped exploring and is only devoted to (robust) exploitation. 

In order to emphasize the \emph{structured} property of the exploration actions, a different system is considered next:
\begin{equation}\label{Nom2}
\begin{array}{l}
A=\left[
\begin{array}{ccc}
  0.9  &  0.5   &      0\\
         0  &  0.9   & 0.2\\
         0 &  -0.2 &   0.8\\
\end{array}
\right]
\end{array},\quad B=\left[
\begin{array}{cc}
  0  &  .1   \\
         0.6  &  0  \\
         0 &  0.6 \\
\end{array}
\right].
\end{equation}
Note that the system has now all its eigenvalues very close to the unit disk, and that the least damped mode is close to become uncontrollable.

\begin{figure}[h!]%
\begin{center}
\includegraphics[width=1\columnwidth]{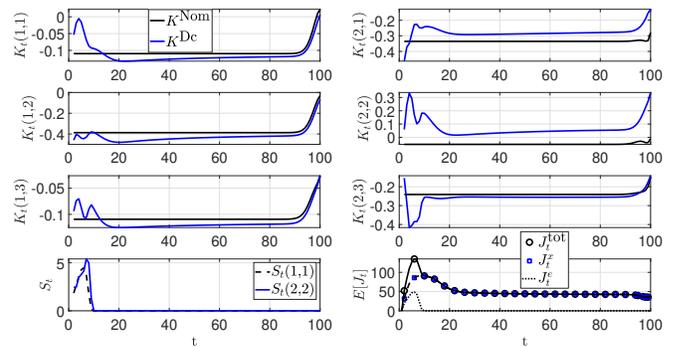}
\caption{Dual controller and cost for system (\ref{Nom2}).}
\label{Rob_Expl_2}
\end{center}
\vspace{-0.1in}
\end{figure}

Coarse-ID estimates of the system are obtained as for (\ref{Nom1}), and the dual control policy is shown in Figure \ref{Rob_Expl_2}.

Exploration actions can again be detected in the first part of the horizon, however this time they are performed by both the state-feedback $K_t$ and the covariance $S_{t}$.
Two types of trade-offs arising in the dual control problem can be appreciated by comparing the different trends in Figs. \ref{Rob_Expl_1}-\ref{Rob_Expl_2}. The first is the one between exploration and exploitation, captured by the fact that the former only lasts for a certain fraction of the total mission.
The second trade-off concerns the choice, for the purpose of exploration, between $K_{t}$ and $S_{t}$.
It is indeed observed, as expected, that whenever it is possible to explore in a \emph{controlled} manner (i.e. without resorting to random excitation), this is the preferred way.
The best exploration strategy
inevitably depends on the true (unknown) plant to control, for example its controllability and margin of stability.
Figs. \ref{Rob_Expl_1}-\ref{Rob_Expl_2} also show that, unlike the nominal case where there is no sensible variation of the optimized controller within the horizon (except for the very last timesteps, recall Fig. \ref{Cmprs_Nom}),
the dual task for which the policy $G_{\text{K}}(K_t,S_{t})$ is designed makes the most important features of the problem (e.g. $K_t$, $S_t$, timestep costs $J_t^{\text{tot}}$) inherently time-varying and thus this type of dual control problem should be studied in a finite horizon setting.

\section{Conclusions}

The paper proposes a dual control synthesis approach for the finite horizon LQ problem. A control law is designed with the
twofold objective of minimizing the worst-case quadratic cost in the face of an ellipsoidal uncertainty set while reducing it based on the system response.
This is achieved by formulating an application-oriented, since the effect of the policy on the ellipsoidal set is captured in the optimization problem,
and safe, since the designed controller is robust, exploration action. SDP programs to solve the nominal, robust (with fixed uncertainty) and dual control problems are proposed, and their application is shown.
The resulting exploration encompasses different types of trade-offs and shows how the optimal actions
depend on the features of the true plant. 

\bibliography{biblioLQR}             
\section*{Appendix}
Proof of Lemma 1.

\begin{proof}
Recall from Section \ref{SDP_formulation} the definitions:
$P_t=\mathbb{E} \left [
x_{t}
x_{t}^{\top} \right]$, $\bar{Q}_t:=\left[ \begin{smallmatrix}Q^{\frac{1}{2}} \\ R^{\frac{1}{2}}K_t \end{smallmatrix}\right]$, $\bar{R}:=\left[ \begin{smallmatrix}0 \\ R^{\frac{1}{2}} \end{smallmatrix}\right]$. Define $M_t=Q+K_t^{\top} R K_t$. By virtue of the chosen policy (\ref{feedback_law}), $J$ can be rewritten as:
\begin{equation}\label{fh_cost_1}
J= \mathbb{E} \left[ \sum_{t=1}^{T-1} \left(x_{t}^{\top} M_t x_{t}+e_{t}^{\top}R e_{t}\right)+x_{T}^{\top}Q x_{T} \right]
\end{equation}
Consider the first term in the summation (i.e. the one that depends on $x_t$). Simple matrix manipulations give it as:
\begin{equation}\label{fh_cost_2}
\begin{aligned}
&\mathbb{E} \left[ \sum_{t=1}^{T-1} x_{t}^{\top} M_t x_{t} \right]=
\mathbb{E} \left[x^{\top} \left( I_{T-1} \otimes M_t \right)x\right]\\
=& \mathbb{E} \left[\textnormal{Tr}\left( x x^{\top} I_{T-1} \otimes M_t \right)\right]=\textnormal{Tr}\left( \mathbb{E} \left[x x^{\top}\right] I_{T-1} \otimes M_t \right)\\
\end{aligned}
\end{equation}
where $x\in \mathbb{R}^{(T-1) n_x}$ denotes the vector obtained stacking the states $x_t$, $\mathcal{W}=\mathbb{E} \left[x x^{\top}\right] \in \mathbb{S}^{(T-1) n_x}$ denotes the covariance matrix of the state over the horizon and $\otimes$ is the Kronecker product.
It follows that:
\begin{equation}\label{fh_cost_4}
\textnormal{Tr}\left(\mathcal{W} I_{T-1} \otimes M_t \right)= \textnormal{Tr}\left(\left(I_{T-1} \otimes \bar{Q}_t \right) \mathcal{W} \left(I_{T-1} \otimes \bar{Q}_t \right)^{\top} \right)
\end{equation}
Due to the block diagonal structure of $\left(I_{T-1} \otimes \bar{Q}_t\right)$, it follows that:
\begin{equation}\label{fh_cost_5}
\textnormal{Tr}\left(\left(I_{T-1} \otimes \bar{Q}_t \right) \mathcal{W} \left(I_{T-1} \otimes \bar{Q}_t \right)^{\top} \right)
=\textnormal{Tr}\left(\sum_{t=1}^{T-1} \left(\bar{Q}_t P_t \bar{Q}_t^{\top} \right)\right)
\end{equation}
The contribution to the cost only depends on the \emph{diagonal} terms of $\mathcal{W}$, which are the covariance matrices $P_t$ at the various timesteps:
\begin{equation}\label{fh_cost_5}
\mathbb{E} \left[ \sum_{t=1}^{T-1} x_{t}^{\top} M_t x_{t} \right]=\textnormal{Tr}\left(\sum_{t=1}^{T-1} \left(\bar{Q}_t P_t \bar{Q}_t^{\top} \right)\right)
\end{equation}
The contribution of the state to the cost at $t=T$
directly follows from (\ref{fh_cost_5}) specializing it  to the case when $u$ is not penalized and thus $\bar{Q}_T\equiv Q$. Therefore:
\begin{equation}\label{fh_cost_6}
\mathbb{E} \left[ Q P_T Q^{\top} \right]=\textnormal{Tr}\left( Q P_T Q^{\top} \right)
\end{equation}
Finally, the proof for the term depending on $e_t$ in the cost (\ref{fh_cost_1}) follows along the same lines. This is further simplified by the fact that $e_t$ are uncorrelated for different times, and thus their covariance matrix in the horizon has the block diagonal structure $\left(I_{T-1}\otimes S_t\right)$.

\end{proof}

\end{document}